Dynamical effects on the classical Kuiper Belt during the excited-Neptune model[*]

Rafael Ribeiro de Sousa,[1,2] Rodney Gomes,[3] Alessandro Morbidelli,[2] and Ernesto Vieira Neto[1]

[1]*São Paulo State University, UNESP,Campus of Guaratinguetá, Av. Dr. Ariberto Pereira da Cunha, 333 - Pedregulho, Guaratinguetá - SP, 12516-410, Brazil*

[2]*Laboratoire Lagrange, UMR7293, Université Côte d'Azur, CNRS, Observatoire de la Côte d'Azur, Boulevard de l'Observatoire, 06304 Nice Cedex 4, France*

[3]*Observatório Nacional, Rua General José Cristino 77, CEP 20921-400, Rio de Janeiro, RJ, Brazil*

## ABSTRACT

The link between the dynamical evolution of the giant planets and the Kuiper Belt orbital structure can provide clues and insight about the dynamical history of the Solar System. The classical region of the Kuiper Belt has two populations (the cold and hot populations) with completely different physical and dynamical properties. These properties have been explained in the framework of a sub-set of the simulations of the *Nice Model*, in which Neptune remained on a low-eccentricity orbit (Neptune's eccentricity is never larger than 0.1) throughout the giant planet instability (Nesvorný 2015a,b). However, recent simulations (Gomes et al. 2018) have showed that the remaining *Nice model* simulations, in which Neptune temporarily acquires a large-eccentricity orbit (larger than 0.1), are also consistent with the preservation of the cold population (inclination smaller than 4 degrees), if the latter formed *in situ*. However, the resulting a cold population showed in many of the simulations eccentricities larger than those observed for the real population. The purpose of this work is to discuss the dynamical effects on the Kuiper belt region due to an excited Neptune phase. We focus on a short period of time, of about six hundred thousand years, which is characterized by Neptune's large eccentricity and smooth migration with a slow precession of Neptune's perihelion. This phase was observed during a full simulation of the *Nice Model* (Gomes et al. 2018) just after the last jump of Neptune's orbit due to an encounter with another planet. We show that if self-gravity is considered in the disk, the precession rate of the particles longitude of perihelion $\varpi$ is slowed down, which in turn speeds up the cycle of $\varpi_N - \varpi$ (the subscript $_N$ referring to Neptune), associated to the particles eccentricity evolution. This, combined with the effect of mutual scattering among the bodies, which spreads all orbital elements, allows some objects to return to low eccentricities. However, we show that if the cold population originally had a small total mass, this effect is negligible. Thus, we conclude that the only possibilities to keep at low eccentricity some cold-population objects during a high-eccentricity phase of Neptune are that (i) either Neptune's precession was rapid, as suggested by Batygin et al. (2011) or (ii) Neptune's slow precession phase was long enough to allow some particles to experience a full secular cycle of $\varpi - \varpi_N$.

Corresponding author: Rafael Ribeiro de Sousa
rafanw72@gmail.com, rribeiro@oac.eu

rodney@on.br

## 1. INTRODUCTION

The Kuiper Belt Objects (KBOs) are a collection of icy bodies located beyond Neptune's orbit. They are classified into five dynamical classes (more details in Gladman et al. (2008)): (i) resonant populations: objects inside Neptune's mean motion resonances; (ii) scat-tering population: objects whose orbits are repeatedly perturbed by close encounters with Neptune; (iii) classical cold population: objects on fairly eccentric orbits and low inclinations; (iv) classical hot population: objects have moderately eccentric orbits but with larger inclinations and (v) detached population: objects on high-eccentricity orbits that were presumably scattered by Neptune in the past, but had their perihelion increased by resonances with Neptune and presently have no close encounters with Neptune.

The orbital eccentricity and inclination distribution of the observed Kuiper Belt Objects is plotted in Figure 1. The five dynamical classes presumably formed due to the gravitational influence of the four major planets during their migration phase. Neptune, being the outermost planet, had a particularly important influence on the sculpting of the trans-Neptunian region. Thus, the dynamical history of Neptune is an essential ingredient to explain the formation of the observed structure of the Kuiper Belt.

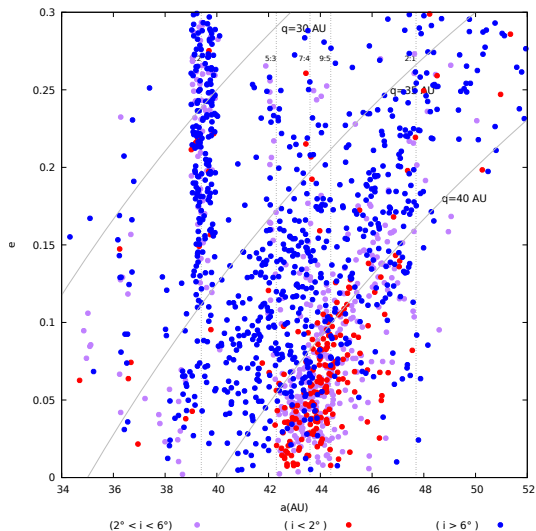

**Figure 1.** The eccentricity distribution (osculating elements) of the observed Kuiper Belt objects from the Minor Planet Center (MPC). We used the KBOs classification of Dawson & Murray-Clay (2012). The red, blue and purple objects have $i < 2°$, $i > 6°$ and $2° \leq i \leq 6°$, respectively. We can see objects in mean motion resonances with Neptune (resonant populations), distributed along vertical lines. Other objects are scattered by Neptune and are distributed in a band bounded by the the pericenter lines $q = 35$ AU and $q = 38$ AU ("scattering" population). The cold and hot classicals are two populations decoupled from resonances and not suffering close encounters with Neptune. Effectively, these populations are confined within $a < 48$ AU.

The classical cold and hot objects are very interesting populations of the Kuiper Belt. They are located between the 3:2 and 2:1 mean motion resonances with Neptune, i.e. in the range of the semi-major axis from 39.5 to 47.8 AU, with eccentricities smaller than 0.24. The classical objects satisfy two general important dynamical features: they have non-resonant orbits and their perihelia are far enough to avoid scattering by Neptune. Despite these similarities, the cold and hot populations have two different inclinations distributions (Brown 2001). We have separated the cold and hot objects following Dawson & Murray-Clay (2012), who avoid a single cutoff in inclinations that would result in a misclassification of some hot and cold objects. They suggest the following inclination classification: a cold population with inclination $i < 2°$, a hot population $i > 6°$ and an ambiguous population ( $2° < i < 6°$).

Dawson & Murray-Clay (2012) showed that the cold population does not fill the entire region that is shown to be stable in the dynamical map of Lykawka & Mukai (2005). In particular, the cold population is confined to eccentricities $e < 0.1$, in the region of semi major axis from 42.5 to 44.5 AU, although stability would be possible also for somewhat larger eccentricities. The clump of objects in this region is known as the "kernel" (Petit et al. 2011). In contrast, the hot population seems to fill the entire stability region. From this observation, they suggested that, while the hot population was probably implanted into the Kuiper belt from smaller heliocentric distances, the cold population is local and was never substantially excited in orbital eccentricity and inclination (however, see Morbidelli et al. (2014) for an alternative explanation). In addition to these different eccentricity and inclination distributions, the cold and hot populations have also different physical properties. The cold classical population has redder colors than the hot population, presumably due to weathered surfaces rich in ammonia ices (Brown et al. 2011; Nesvorný 2015a) moreover their albedo are generally larger than those of the hot population (Brucker et al. 2009). The cold classical population has a size distribution with a steeper slope for $D > 100$ km objects and also lacks very large objects (Bernstein et al. 2004; Fraser et al. 2014; Levison & Stern 2001). Curiously, the hot classical population has colors, albedo, and size distribution similar to those of the other dynamical classes (resonant, scattered and detached).

The discovery of the Kuiper Belt has provided new constraints for Solar System formation models. In general, it is well accepted by the scientific community that the giant planets migrated from their initial location after gas dissipation, due to their interaction with the remaining planetesimal disk (Fernandez & Ip 1984). Another important point of consensus is that the original trans-Neptunian disk was much more massive than the current KBO population, and it was substantially depleted and sculpted by the evolution of Neptune. The only exception would be represented by the cold population, which formed *in situ* and was never substantially affected by Neptune's evolution. The wide binaries, which are common in the cold population, would have been efficiently destroyed by scattering of Neptune. This indicates that the cold population should have formed in situ (Parker & Kavelaars 2010). The *in situ* formation of the cold population would also explain the physical differences with the other classes of the KB, which would be made of objects transported to their current orbits from originally smaller heliocentric distances during the evolution of Neptune's orbit.

The first hint for a big change of Neptune's orbit came from of the discovery of the resonant population. Malhotra (1993, 1995) suggested that Neptune migrated outwards on a nearly circular orbit. The resonance locations slowly moved with Neptune, capturing objects from the trans-Neptunian disk. The radial migration of Neptune on a circular orbit would also explain the origin of the hot population, as a collection of objects scattered by Neptune and then trapped on large-$q$ orbits (see Gomes (2003)). However, several features of the Solar System (e.g. the non-negligible eccentricities of the planets) suggest that the orbital evolution of the planets was not as simple as a smooth radial migration.

Tsiganis et al. (2005), Gomes et al. (2005) and Morbidelli et al. (2005) introduced the so-called *Nice Model*. The *Nice Model* is a scenario that attempted to reproduce the global architecture of the current Solar System by coupling planetesimal-driven migration with a phase of dynamical instability of the planets. The most up-to-date version of the *Nice Model* starts with five giant planets and a massive planetesimal trans-Neptunian disk. The planetary system was originally in a compact, fully resonant configuration, consistent with the phase of radial migration of the planets in the gas-disk (Morbidelli et al. 2007). After the removal of the gas, under the effects of the planetesimals, the planets escaped from their multi-resonant configuration and became unstable. The dynamical evolution of the planets then proceeded due to a combination of mutual close encounters and planetesimal driven migration, until reaching their current orbits (Nesvorný & Morbidelli 2012).

Nowadays, two types of Neptune's dynamical evolutions have been suggested to explain the Kuiper Belt: the excited-Neptune evolution and quasi-smooth Neptune's migration. In the excited-Neptune evolution, Neptune suffered close encounters with the other planets and had a transient phase with an eccentric orbit ($e > 0.1$). The high eccentric phase generated a chaotic sea between the 3:2 and 2:1 MMRs, allowing planetesimals to be captured in this region (Levison et al. 2008). However, Dawson and Murray Clay (2012) investigated the properties of the chaotic sea in the classical region and concluded that the existence of the chaotic region depends of the details of Neptune's interaction with Uranus. The excited-Neptune evolution could explain the hot, resonant and scattered populations. Recent simulations by Gomes et al. (2018) showed that the excited-Neptune evolution is also consistent with the existence of the cold population in terms of inclination distribution, if the latter formed *in situ*. Earlier simulations of Barucci et al. (2008) and Batygin et al. (2011) also demonstrate the survival of locally generated cold KBOs in the high-eccentricity Nice model.

The quasi-smooth Neptune's migration was proposed by Nesvorný (2015a,b) to explain two features of the Kuiper Belt: the observed inclination distribution of the hot population and the "kernel" of the cold population. Nesvorný (2015a) argued that Neptune migrated through the disk on an e-folding timescale $\tau \geq 10$ My before that the giant planets became unstable. With the slow migration of Neptune there was enough time for dynamical processes to raise the inclinations of the planetesimals and reproduce quantitatively the inclination distributions of the hot classical and resonant objects. The smooth migration of Neptune was interrupted by a discontinuous change of Neptune's semi major axis when the giant planets became unstable. If this happened when Neptune was at 28 AU, the Kuiper belt "kernel" would be explained as the collection of objects captured into the 2:1 MMR with Neptune and transported outwards during Neptune's smooth migration phase, then released from resonance during Neptune's semi major axis jump (Nesvorný 2015b). During the instability, Neptune attains a non-negligible eccentricity which distinguishes this model from the smooth migration model proposed by Malhotra (1993, 1995) where Neptune's migration retains always a circular orbits.

Although the excited-Neptune scenario is able to reproduce in some cases the distribution of eccentricities of the cold belt (Gomes et al. 2018), there are many cases in which there is a lack of low-eccentricity objects. Batygin et al. (2011) showed that the slow precession of Neptune during its eccentric phase is responsible for exciting the eccentricities of the cold belt. They showed with a simple analytical model that, if Neptune's eccentric phase had been characterized by a fast precession of the perihelion, the cold population would have preserved an unexcited state in eccentricity. Unfortunately, however, the slow precession of Neptune typically occurs, at least temporarily, during the planet's high-eccentricity phase.

The purpose of this work is to discuss the dynamical effects happening in the classical region during the excited-Neptune phase. We focus on a short period of time, of about six hundred thousand of years, that is characterized by a slow Neptune's precession, a large eccentricity and a smooth migration. This phase was observed during a full simulation of the *Nice model* (Gomes et al. 2018) just after the last jump of Neptune's orbit due to an encounter with another planet. It is described in Section 2. Because of the eccentricity acquired by Neptune's orbit during the jump, the cold population had been already excited in eccentricity before the beginning of this phase. In section 3, we show that, if the mutual interactions among the planetesimals are taken into account (self-gravity) and there is substantial mass in the planetesimal disk, the precession of the perihelia relative to Neptune's (i.e. of the angle $\varpi - \varpi_N$) becomes faster and therefore the phase corresponding to low eccentricities can be reached. In addition, mutual scattering among the bodies spreads the eccentricity distribution, also helping some objects to reach very low eccentricity values. We also study the more realistic case where the massive planetesimal disk ends at about 30-35 AU and the cold population has a negligible mass from the beginning. In Section 4 we summarize our results and conclude discussing the scenarios that could lead to low-eccentricity objects in the cold population despite of the temporary high-eccentricity phase of Neptune.

We warn the reader that the aim here is not reconstructing quantitatively the structure of the Kuiper Belt, but to provide a proof of concept that the excited-Neptune evolution, under some conditions, can be consistent with the small eccentricities observed in the cold Kuiper belt population.

## 2. EXCITED-NEPTUNE EVOLUTION

The evolution of the giant planets orbits are taken from the Nice Model simulations performed by Gomes et al. (2018). This simulation satisfies the "success" criteria described in Nesvorný & Morbidelli (2012) (Gomes et al. 2018). It starts with five planets (Jupiter, Saturn and three Neptune-mass planets) all in mean motion resonances with each other, as expected from giant planet migration in a protoplanetary disk of gas (Morbidelli et al. 2007). Unlike Nesvorný (2015b), Gomes et al. (2018) chose a compact multi-resonant configuration ( 3:2, 3:2, 4:3 and 5:4 ) and thus they obtained a more violent instability (Figure 2). The three outermost giant planets have $4.5 \times 10^{-5}$ solar masses and Neptune is just defined as the outermost planet at the end of the integration and the fifth planet is the one that is ejected.

In addition to the planets, the system comprises a disk of planetesimals, located between the initial location of the outermost planet and 45 AU. The initial surface density of the disk is $\Sigma(r) \propto 1/a$ and the disk is modeled as a collection of 4,000 equal-mass bodies, with a total mass of the planetesimal disk $M_{disk} = 35 M_{Earth}$; thus each particle has 4 Pluto's mass. The eccentricities were chosen randomly between 0 and 0.002. The inclinations were initially null and the other angles were chosen randomly between 0 and 360 degrees. The mutual gravitational interaction among the particles are neglected. In the simulation Neptune has a jump in semi-major axis from its local formation to 24 AU as a consequence of close encounters with the other planets. With the last jump, just before 33 My, Neptune reaches an eccentricity of $\sim 0.27$. Then, the eccentricities of the planets decrease due to dynamical friction with the planetesimal disk and a slow-smooth migration phase takes place. This final phase continues until the planetary system reaches approximately the current semi major axes, eccentricities and inclinations.

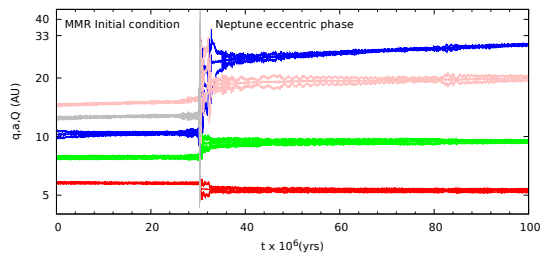

**Figure 2.** Orbital evolution of the semi major axes, aphelion and perihelion distances of the planets for one of the successful Nice-model simulations of Gomes et al. (2018). The planetary system starts from a compact multi-resonant configuration ( 3:2, 3:2, 4:3 and 5:4 MMR ) and undergoes a temporary period of instability during which the filth giant planet (gray) is ejected by a close encounter with Jupiter (red). Neptune (blue) undergoes for a short period of high eccentricity after its encounters with Saturn (green) and Uranus (pink). The system ultimately evolves towards the current semi major axes and eccentricities of the real giant planets of the Solar System.

The state of the particle disk at 33 My is shown in Figure 3. As in Fig. 1, blue, magenta and red dots denote the particles with $i > 6°$, $2° \leq i \leq 6°$ and $i < 2°$ respectively. The black dots show the real members of the cold population. As one sees from the top panel of the figure, already at the beginning of the high-eccentricity phase of Neptune, the synthetic cold population (red dots) is too excited in eccentricity compared to the real one. The excitation occurred during Neptune's encounter phase. Thus, there is a clear lack of bodies with eccentricities smaller than 0.05.

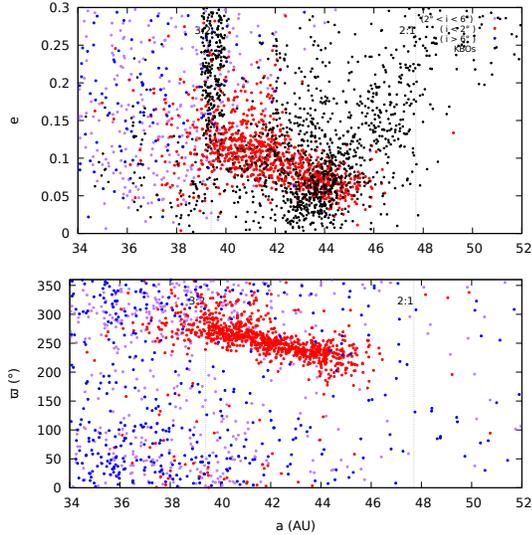

**Figure 3.** The eccentricity (top panel) and perihelion longitudes (bottom panel) as a function of the semi-major axis for the trans-Neptunian planetesimal disk at 33 Myr in the simulation of Fig. 2. The red, blue and purple objects have inclinations of $i < 2°$, $i > 6°$ and $2° \leq i \leq 6°$, respectively. The black points show the observed Kuiper Belt Objects. The planetesimal disk already shows the early structure of the Kuiper Belt, with the scattered, resonant, hot and cold populations. However, the cold population is clustered in perihelion longitude and has a deficit of $e < 0.05$ orbits compared to the observed population.

The goal of this paper is to study how this result could change and whether some particles could remain/return to a quasi-circular orbit if the self-gravity of the disk were taken into account. For this goal, we need to redo the simulations without neglecting the gravitational interactions among the particles. However, because the system is strongly chaotic, each simulation would produce a radically different planetary evolution, in most cases incompatible with the current state of the Solar System. In fact, out of 2,000 numerical simulations, Gomes et al. (2018) obtained only 53 good ones in terms of the final planetary orbits.

To avoid this problem, we fix the planetary evolution to that shown in Fig. 2 in order to investigate the evolutions of the planetesimals always in the same planetary perturbation framework. This is done by interpolating the orbital elements of the planets from the output of the Gomes et al. simulation using spline functions and then using this interpolated evolution in the new simulations. This strategy has been already used in a number of studies (e.g. Morbidelli et al. (2009); Brasser et al. (2009); Nesvorný (2011)).

The interpolated evolutions of the semi major axes, eccentricities, inclinations, longitudes of the perihelion ($\varpi$), node ($\Omega$) and mean ($\lambda$) longitudes of the planets are showed in Figure 4. For the interpolation, we used cubic splines. We interpolated all orbital elements between two successive outputs. The time-resolution of the output in Gomes et al. 2018 simulation was 100,000y. Because this exceeds the orbital period of a body, we calculated the number of orbits between two successive outputs using the information on the mean orbital period (from the mean semi major axis) and then adjusted the frequency so that the value of $\lambda$ at the end of each output timestep coincided with that recorded in the simulation (Figure 4 (f)). Batygin et al. (2011) defined the fast and slow precession of Neptune's perihelion in terms of the secular frequency $g_8$ (the current frequency of the perihelion for Neptune). We focus on the phase between 33 and 33.6 My, because the eccentricity of Neptune is large (from 0.27 to 0.20) and the precession of its longitude of perihelion is slow ($g_{Neptune} < g8$). The high-eccentricity, slow precession phase is supposedly the most dangerous one for the excitation of the cold population (Batygin et al. 2011). Thus, if we will be able to obtain low-eccentricity particles during this phase it is likely that a cold population can be preserved throughout the whole evolution. We stress that self-gravity simulations are very slow and therefore it would not be possible to cover the whole time-range of Fig. 2. Thus, we need to select an interesting (and a priori the most defavorable) interval of time and restrict our numerical analysis on this interval. We adapted the integration package REBOUND (Rein & Spiegel 2015) to read the interpolated evolutions of the planets shown in Fig. 4 instead of solving for the planetary motion self-consistently. The evolution of the particles is computed from the gravitational forces exerted by the planets from their interpolated positions.

It is important to realize (bottom panel of Fig. 3) that the planetesimals with inclinations smaller than 2°, and $a > 42$ AU (red dots), have clustered perihelion longitudes, whereas the particles with large inclinations (blue dots) have perihelion longitudes much more randomized. This suggests that the cold population, which was originally *in situ* on $e \sim 0$ orbits, has been excited by secular perturbations from Neptune (so that all particles had a coherent evolution in the $e, \varpi$ plane), whereas the objects of the hot population, which came predominantly from smaller semi major axes, have suffered close encounters with the planet and/or mean motion resonant trapping, which dispersed their perihelion longitudes.

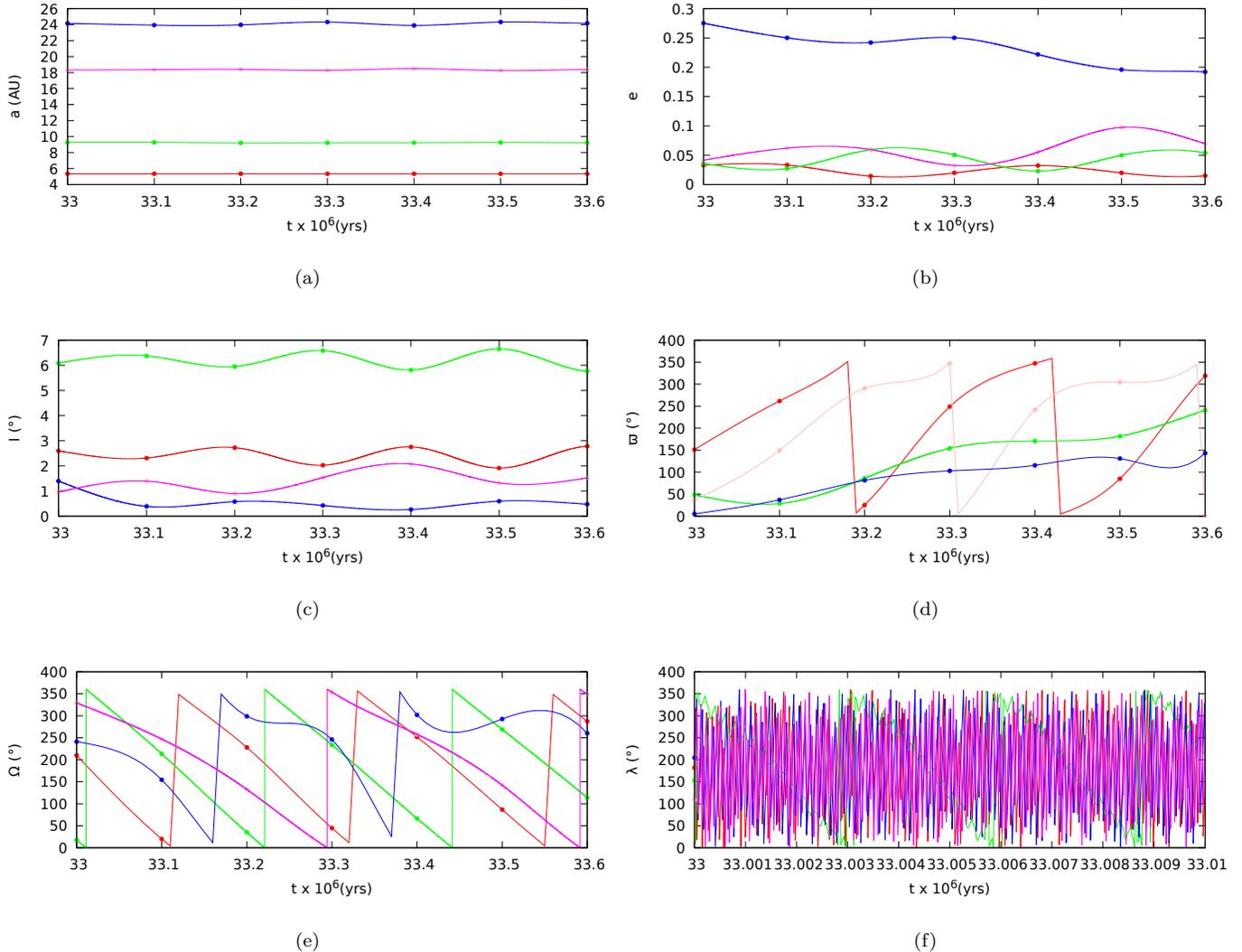

**Figure 4.** The orbital evolution of the giant planets during Neptune's high eccentricity phase. The points represent the orbital elements of Jupiter (red), Saturn (green), Neptune (blue) and Uranus (pink) recorded in the output of the simulation of Fig. 2 from 33 to 36.6 My. The curves represent the synthetic dynamical evolution of the giant planets produced by the spline interpolation of the output points. Neptune decreases its eccentricity from 0.27 to ~ 0.20 during 0.6 My. The semimajor axis of Neptune does not change much during this short period of time and the longitude of perihelion has a very slow precession.

## 3. RESULTS

In this section we investigate whether the self-gravity among planetesimals could produce objects with eccentricities as small as those of the real cold population. For this goal, we adopted the planetary evolution presented on section 2 and performed N-body simulations of the particles evolution accounting for the self-gravity of the disk. We remind the reader that the phase of Neptune's evolution we focus on (slow Neptune's precession and a very eccentric orbit) is a priori the least favorable to obtain particles on small-eccentricity orbits.

Aiming at measuring for which planetesimal masses the self-gravity effect is important we performed three simulations in which the individual planetesimals had 4, 1 and 0.01 Pluto's masses, respectively. Because the total number of planetesimals in the disk is given (4,000 in the original Gomes et al. (2018) simulation), a smaller individual planetesimal mass corresponds to a smaller disk mass. The mass of the Kuiper Belt estimated from (Fraser et al. 2014) and CFEP-OSSOS (Bannister et al. 2018) are $3 \times 10^{-4}$ and 0.01 Earth's mass. Our simulation started at 33 My with 706 objects inside of the cold population ($a > 42$ AU, $i < 2$ degrees and $e < 0.1$) considering the individual planetesimal mass had 4 Pluto's masses, 1 Pluto's masses and 0.01 Pluto's masses the total mass of the cold population is aproximately 6.0, 1.5 and 0.015 Earth masses, respectively. In the end of the simulation at 33.6 My, the numbers of the objects in the cold population for the cases with 4, 1 and 0.01 Pluto's mass are 339 objects, 344 objects and 246 objects, respectively. It correspond to final masses of 2.85, 0.72, 0.005 Earth's mass for each case of our simulations.

As anticipated in the previous section, our simulations start from the state of the system observed in Gomes et al. (2018)'s simulation at 33 My and cover a 0.6 My timespan (after which Neptune's precession becomes fast). In a first set of simulations, we consider "massive" only the planetesimals located in the region of semimajor axis from 30 to 60 AU at $t = 33$ My. This region includes the cold population, the hot population and a part of scattered disk. This set of simulations is interesting to understand the role of self-gravity in general but, as anticipated in the introduction and detailed more below, it is not realistic because the cold belt presumably never contained a substantial mass. In a second set of simulations, we will simulate the effects of a massive scattered disk on a mass-less cold population.

Figure 5 shows the eccentricities as a function of semi major axis of the planetesimals at the end of our simulations. The cases with no-self gravity and the case where planetesimals have 4 Pluto's masses are showed in Fig. 5 (a) and (b), respectively. The blue points depict the particles of the planetesimal disk, the red points represent the observed objects of the Kuiper Belt. Comparing both cases, we observed that the self-gravity among 4 Pluto's mass objects produced a significant number of objects with eccentricity smaller than 0.1, compatible with the observed cold population.

The self-gravity cases with 1 Pluto's mass and 0.01 Pluto's mass are showed in Fig. 5 (c) and (d), respectively. For the case with 1 Pluto's mass, we have a dispersion in eccentricities similar to that observed in the case of 4 Pluto's mass planetesimals, but with fewer objects with eccentricities smaller than 0.1. We observed that the dispersion in eccentricity decreased dramatically for the case with objects of 0.01 Pluto's mass, for which the final distribution is similar to that of the no self-gravity case.

It is well known that the current cold population contains very little mass (Fraser et al. 2014; Bannister et al. 2018). The simulations of Gomes et al. (2018), as well as those of Nesvorný (2015a,b) show that this population should have lost less than 90% of its bodies during the migration and instability of the giant planets. This implies that the disk located beyond $\sim 40$ AU was never significantly massive (probably less than one Mars-mass in total). Instead, the hot population was implanted from within 30-35 AU and represents just a very small proportion ($\sim 0.1\%$) of the original population in that part of the disk (Gomes 2003; Nesvorný 2015a). This is consistent with a disk's mass within 30 AU of 20-30 Earth masses, also required to drive the dynamical evolution of the *Nice model*.

Given these considerations we investigated whether the influence of the collective gravity of the planetesimals coming from within 30 AU could generate low-eccentricity orbits in a massless cold population. For this purpose, we have done a simulation with the planetesimal disk split into two parts: the particles with inclination smaller than 5° at $t = 33$ My are considered to be part of the cold population and are treated as massless particles: instead the hot and scattered populations, which come from the inner part of the disk, are assumed to be made of 1 Pluto's mass planetesimal. We then analyzed the possibility of obtaining the dispersion in eccentricity necessary to keep some particles of the cold population with low enough eccentricities to be consistent with the observed population. The result of this simulation is shown in the Fig. 5 (e).

We observed almost no dispersion effect on the eccentricity distribution of the cold population, obtaining a result very similar to that of the no self-gravity case. We can thus infer that the perturbation of the hot population on the cold population is not significant to provide a dispersion towards low eccentricities of the cold population.

To confirm that the results illustrated in the Fig. 5 (b) and (c) were mostly due to the self-gravity of the cold population on itself, we did another experiment where we assumed that the hot and scattered populations are massless and the cold population is made of 1 Pluto's mass planetesimals. The result of this simulation is showed in Fig. 5 (f). As expected, we observe a dispersion to low eccentricities of the cold population, that resembles the observed population better. Nevertheless, the number of objects reaching low eccentricities is smaller than for the case illustrated in Fig. 5 (c), where both hot and cold population particles had been assumed to have 1 Pluto's mass.

We now proceed to analyze and explain the results presented above. Our analysis is made in in the framework of the secular perturbation theory of free and forced elements (Murray & Dermott 2000). In Figure 6, we plot the evolution of the particles in the $h = e \cos(\varpi - \varpi_N)$ versus $k = e \sin(\varpi - \varpi_N)$ plane, for each case simulated in this section. The colors indicate the time at which particles are plotted, following the scale depicted on the right hand-side bars. The evolution of the particles goes from black bar color ($t = 33.0$ My) to red color ($t = 33.6$ My). The blue and cyan

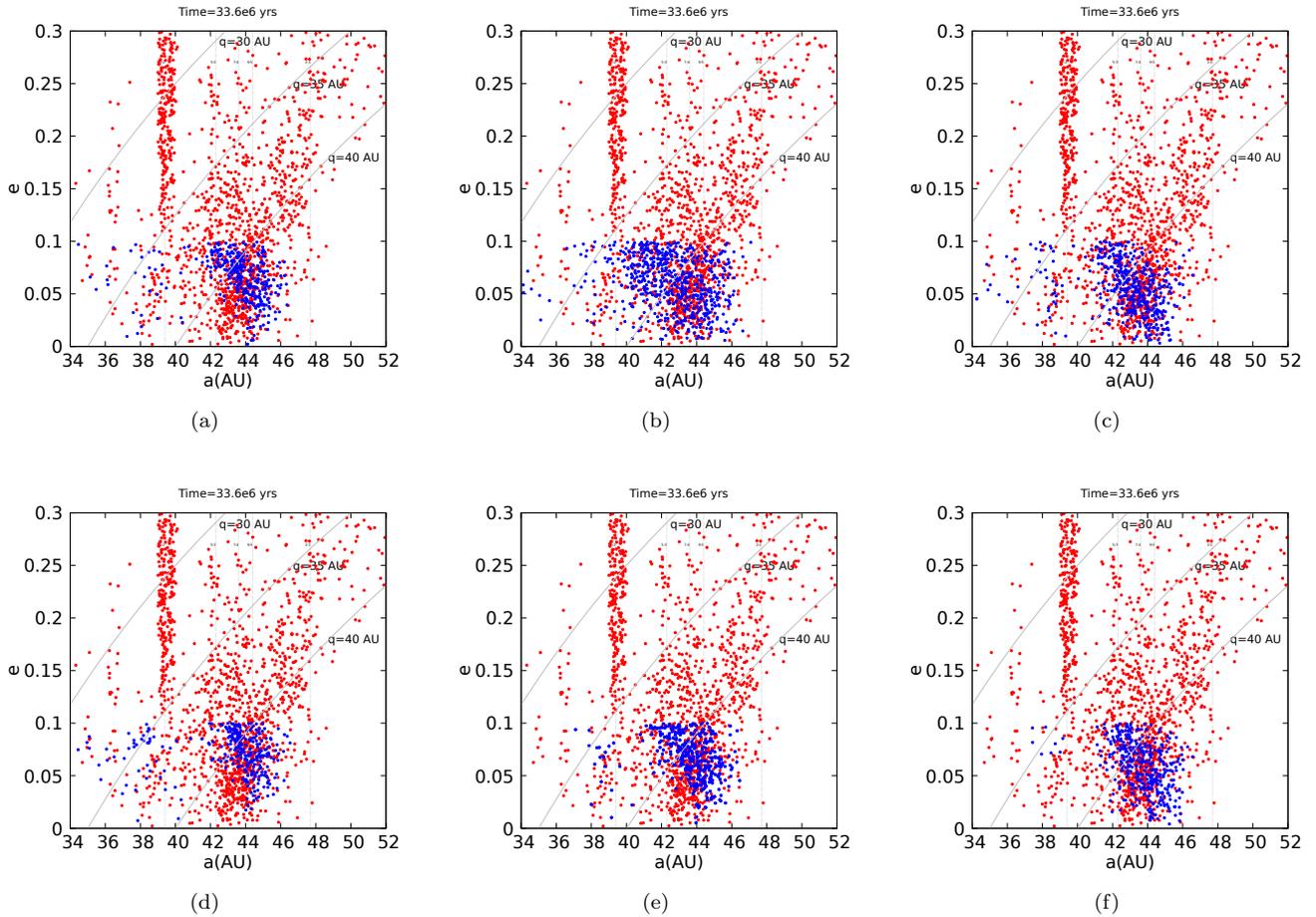

**Figure 5.** The panels show the final eccentricity distribution (at $t = 33.6$ My) of the cold population in simulations assuming different planetesimals masses. The (a) panel refers to a control simulation without self-gravity. The (b) panel is for planetesimals with individual masses of 4 Pluto's mass, the (c) panel has planetesimals of 1 Pluto's mass and the (d) panel 0.01 Pluto's mass. The simulation of the influence of 1 Pluto's mass hot population on a massless cold population is shown in the (e) panel. The simulation with the self-gravity only among planetesimals in the cold population (each with 1 Pluto's mass) is shown in (f) panel. For these plots, we removed particles with $e > 0.1$ and inclinations larger than 4 degrees. The red points represent the observed distribution of the cold Kuiper Belt objects and the blue points depict the simulated objects.

points represent the initial and final distribution of the particles in the h-k plane.

In the no self-gravity case (Fig. 6 (a)), we observe that the evolution traces clockwise arcs of circles in the h and k plane and the center of these arcs is roughly at approximately $(-0.025, 0)$. This point corresponds to the forced eccentricity vector induced by the eccentric Neptune. It shows that the evolution of the particles is dominated by the secular interaction with Neptune. Given the initial phase $\varpi - \varpi_N \sim 120°$, the particles reach $\varpi - \varpi_N \sim 180°$ at the end of the simulation, i.e. the maximum of their forced eccentricity cycle.

However, for the complete (all the disk is self-gravitating) self-gravity case with 4 and 1 Pluto's masses (Fig. 6 (b) and (c) ) we observe two effects. First, it is well-known that the effect of the self-gravity in a disk is to slow down the precession frequency of $\varpi$ (Binney & Tremaine 2008). Consequently, $\varpi - \varpi_N$ precesses clockwise *faster* [1]. In fact, we see in Fig. 6 (b) that the particles reach on average $\varpi - \varpi_N = -45°$. This brings the particles towards the minimum of their eccentricity cycle, which happens at $\varpi - \varpi_N = 0°$. Second, particles have close encounters with each other, which forces all orbital elements to diffuse. Thus, the particles spread on the $(h, k)$ plane, which helps having some particles near the origin at the final time. In Fig. 6 (c) the diffusion process seems to dominate over the precession process, whereas in panel (b) it is the opposite.

---

[1] Although $\varpi - \varpi_N$ becomes a faster angle, the evolution should not be confused with that proposed in (Batygin et al. 2011). In their scenario, $\varpi - \varpi_N$ has to circulate so fast that the forced eccentricity felt by the particles is very small. In our case, $\varpi - \varpi_N$ is not circulating fast enough for this behavior. So, the forced eccentricity felt by the particles is still quite large. But $\varpi - \varpi_N$ is now fast enough that, during the time-window of slow Neptune's precession (0.6 My), the particles can experience a full circulation, coming back close to the initial $e \sim 0$ value.

The cases with 0.01 Pluto's mass (Fig. 6 (d)) and with Pluto-mass particles only in the hot disk (Fig. 6 (e)) show that the cold population particles evolve in $\varpi - \varpi_N$ only by 45°. Thus, we have roughly the same dynamical cycles of the no-self gravity case. This is because particles are too small (case d) or because the cold particles are no longer embedded in a massive disk and close encounters with the hot population are inefficient due to the large relative velocities (case e). The case (Fig. 6 (f)) with massive planetesimals only in the cold population shows an evolution very similar to the 1 Pluto's mass case of panel (c).

## 4. DISCUSSIONS AND CONCLUSIONS

This work concerns the dynamical effects that took place in the classical region of the Kuiper Belt, during the early evolution of Neptune. We chose a specific evolution of Neptune in the framework of the *Nice model* for the Solar System evolution, from Gomes et al. (2018). In this evolution five giants planets, initially locked in a multi-resonant configuration, evolve to instability after a violent close encounter between a Neptune-mass planet and Jupiter resulting in the ejection of the smaller planet. During the instability, Neptune is scattered for a brief moment to high eccentricity orbit and induces changes on the planetesimal disk. The period of Neptune's eccentric phase generates a chaotic sea allowing the hot population to be captured in the classical Kuiper belt region. The cold population, assumed to have been formed *in situ*, is partially depleted and excited in eccentricity. At the end of the considered simulation, the cold population (i.e. that particles that remain with $i < 4$ degrees) does not have eccentricities low enough to resemble the current cold classical population. We observed that Neptune's eccentric phase has a slow precession that allows the excitation of the particles eccentricities in the classical region, in agreement with the results of Batygin et al. (2011).

We investigated how the self-gravity of the planetesimal disk could change this picture. For this study, we interpolated the orbits of the giant planets during the brief period of high-eccentricity and slow-precession of Neptune, from 33 to 33.6 My, using a spline functions algorithm. We started our simulations from the orbital distribution of the planetesimal disk observed at $t = 33$ My. This distribution is characterized by a clustering in perihelion longitude of the synthetic cold population (defined as particles with $i < 2°$ and $a > 42$ AU), revealing that this population suffered only the secular effect of Neptune. The consequence of the secular effect of Neptune is a coherent evolution in the h-k plane ($h = e \cos(\varpi - \varpi_N)$, $k = e \sin(\varpi - \varpi_N)$) allowing correlated changes in the eccentricities and longitudes of perihelion. If Neptune has a large orbital eccentricity and a slow precession, the planetesimal disk receives an excessive excitation from the secular dynamics and cannot preserve enough low- eccentricity orbits to be comparable to the real cold population.

Following the secular theory, we showed that the evolution of the classical cold particles in the h-k space follows clockwise arcs of circles displaced by approximately 0.025 units from the origin. Along these arcs, the particles may evolve towards larger or smaller eccentricities depending on their initial phase of $\varpi - \varpi_N$ and the amplitude of the $\varpi - \varpi_N$ evolution (i.e. the length of the arc). Without self-gravity, the particles evolved to the maximum of their secular eccentricity cycle, reaching mostly eccentricities larger than 0.05. On the other hand, if the high eccentric phase of Neptune and the slow precession take a longer time the eccentricities would evolve to low eccentricities following the secular cycles. Therefore, the ideal case for Neptune's evolution to produce the cold population without any other mechanism is the synchronism between the secular cycles of the planetesimals and the duration of Neptune's eccentric, slowly precessing phase. In the case of disk's self-gravity, the precession of $\varpi$ is slowed down, and therefore the range of clockwise excursion of $\varpi - \varpi_N$ during the simulation timescale is increased. Thus, planetesimals can reach values of $\varpi - \varpi_N$ close to 0°, corresponding to the minimum of their eccentricity cycle. In addition, close encounters among particles, spread them on the $(h, k)$ plane. As a combination of these two effects, some particles can be found at very small eccentricity in the end, compatible with the small eccentricities observed in the real cold population.

In our simulations we have verified that this processes work if the disk was composed by large objects, with either 4 or 1 Pluto's mass. For smaller planetesimals masses, 0.01 Pluto's mass- and hence smaller disk masses given that the number of planetesimals is fixed- the effects of self-gravity are negligible. It is possible that a large number of small planetesimals, carrying a large total mass, could provide similar effects to those provided by a population of 4,000 4-Pluto's mass objects, but we could not test this possibility because it is computationally very demanding

It is generally accepted that the original trans-Neptunian disk contained thousands of Pluto-size ob-jects (Stern 2001; Nesvorný & Vokrouhlický 2016). But these objects are expected to have been initially in the inner part of the disk, within 30 AU, and not in the region of the cold population. Also, while the disk within 30 AU is supposed to have contained 20 to 30 Earth masses to drive the dynamics of the *Nice model* (Nesvorný & Morbidelli 2012) and to be a sufficient source of the hot population (Gomes 2003; Nesvorný 2015a,b) the scattered disk and the Oort cloud (Brasser & Morbidelli 2013; Nesvorný 2015a,b), the cold population is supposed to have formed in situ and to have retained at least 10% of its initial mass. Given that the current mass of the cold population is very small (Fraser et al. 2014; Bannister et al. 2018), it is likely that the cold population never contained more than 0.1 Earth's mass altogether.

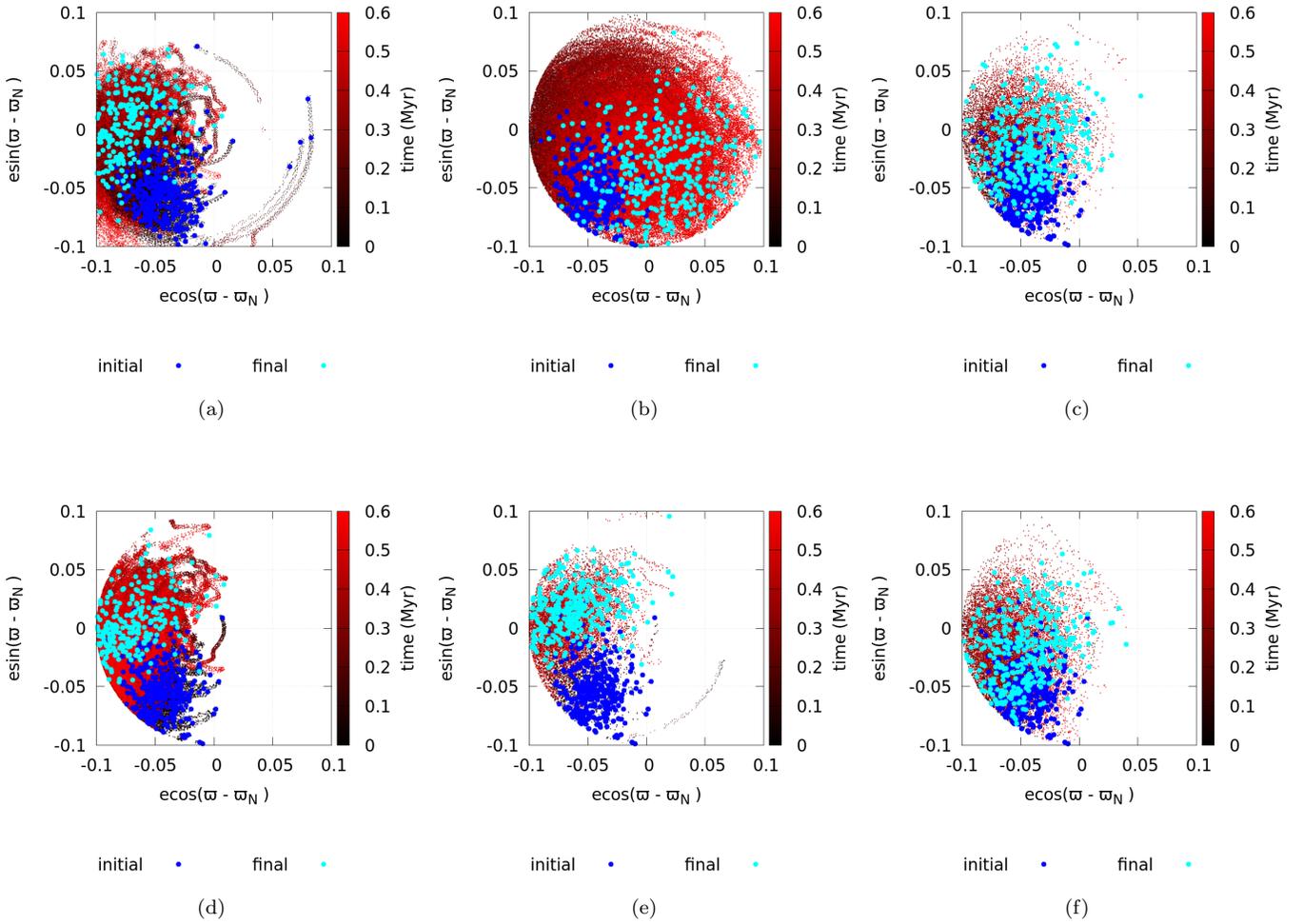

**Figure 6.** The evolution of the particles in the $h = e\cos(\varpi - \varpi_N)$ and $k = e\sin(\varpi - \varpi_N)$ space. The color bars represent the time. The blue and cyan points represent the initial and final distribution of the particles in the h-k plane.

Thus, we have tested whether the sole gravity of the particles dispersed from the inner part of the disk could have effects on the cold population similar to those of the case where the full disk is massive and self-gravitating. Unfortunately, the result is negative. This is probably due to the fact that the cold population is not fully embedded in the hot population, and therefore the modification of the precession rate of $\varpi$ of the cold population is minimal. As for scattering and diffusion in the $(h, k)$ plane, the close encounters between members of the hot population with those of the cold-population happen at high relative velocity and therefore also have a limited effect.

At the light of these results, should we conclude that the small eccentricities observed in the real cold population demonstrate that Neptune never experienced a large-eccentricity phase of any meaningful duration during the giant planet instability? Probably not. As showed by Batygin et al. (2011), if Neptune's high eccentricity is accompanied by a fast precession rate, the small eccentricities of the cold population can be preserved because the forced eccentricity vector dominating the particle's secular evolution is shrunk. Moreover, as we have shown in this paper, even if Neptune's precession is slow and the forced eccentricity vector is non-negligible, the particles eccentricities would evolve to low values following a secular cycle, if the high-eccentricity, slow-precession phase of Neptune lasts long enough. Therefore, the ideal case for Neptune's evolution to produce the cold population without any additional mechanisms is the synchronism between the secular cycles of the planetesimals and the duration of Neptune's eccentric and slow-precession phase. The likelihood that such a synchronism occurs remains to be evaluated but it is expected to be not very large. Thus we conclude that either Neptune experienced a moderately high eccentric phase during its migration (Gomes et al. 2018) or experienced a quasi-smooth migration with a jump at 28 AU (Nesvorný 2015a,b).

Our goal in this paper has been to provide a proof of concept and not to reproduce quantitatively the formation of the structure of the cold classical population. Thus, we limited ourselves to showing in which circumstances the excited-Neptune model could be compatible with the small eccentricities of the cold Kuiper belt population.

## ACKNOWLEDGMENTS

R.R.S acknowledges support provided by grants #2017/09919-8 and #2015/15588-9, São Paulo Research Foundation (FAPESP) and the good hospitality received from Observatoire de la Côte d'Azur. R.R.S also thanks Rafael Sfair for turning possible our simulations on the cluster of computers that belongs to São Paulo State University.

12  Ribeiro et al.Nesvorný, D. 2011, ApJL, 742, L22, doi: 10.1088/2041-8205/742/2/L22

—. 2015a, AJ, 150, 73, doi: 10.1088/0004-6256/150/3/73

—. 2015b, AJ, 150, 68, doi: 10.1088/0004-6256/150/3/68

Nesvorný, D., & Morbidelli, A. 2012, AJ, 144, 117, doi: 10.1088/0004-6256/144/4/117

Nesvorný, D., & Vokrouhlický, D. 2016, ApJ, 825, 94, doi: 10.3847/0004-637X/825/2/94

Parker, A. H., & Kavelaars, J. J. 2010, ApJL, 722, L204, doi: 10.1088/2041-8205/722/2/L204

Petit, J.-M., Gladman, B., Kavelaars, J. J., Jones, R. L., & Parker, J. 2011, 722

Rein, H., & Spiegel, D. S. 2015, MNRAS, 446, 1424, doi: 10.1093/mnras/stu2164

Stern, S. A. 2001, in Bulletin of the American Astronomical Society, Vol. 33, AAS/Division for Planetary Sciences Meeting Abstracts #33, 1033

Tsiganis, K., Gomes, R., Morbidelli, A., & Levison, H. F. 2005, Nature, 435, 459, doi: 10.1038/nature03539